\documentclass[aps,pre,twocolumn,showpacs]{revtex4}
\usepackage{epsfig}
\usepackage{times}
\begin{document}

\title{Self-organizing patterns maintained by competing associations \\
in a six-species predator-prey model}

\author{Gy{\"o}rgy Szab{\'o}, Attila Szolnoki, and Istv{\'a}n Borsos}
\affiliation
{Research Institute for Technical Physics and Materials Science
P.O. Box 49, H-1525 Budapest, Hungary}

\date{\today} 

\begin{abstract}
Formation and competition of associations are studied in a six-species ecological model where each species has two predators and two prey. Each site of a square lattice is occupied by an individual belonging to one of the six species. The evolution of the spatial distribution of species is governed by iterated invasions between the neighboring predator-prey pairs with species specific rates and by site exchange between the neutral pairs with a probability $X$. This dynamical rule yields the formation of five associations composed of two or three species with proper spatiotemporal patterns. For large $X$ a cyclic dominance can occur between the three two-species associations whereas one of the two three-species associations prevails in the whole system for low values of $X$ in the final state. Within an intermediate range of $X$ all the five associations coexist due to the fact that cyclic invasions between the two-species associations reduce their resistance temporarily against the invasion of three-species associations.
\end{abstract}


\maketitle

\section{INTRODUCTION}
\label{sec:intro}

Many real systems consist of small different objects whose organization into large spatial associations (communities) is the result of some evolutionary rules controlling the system's behavior at the microscopic level \cite{watt_je47,gilpin_an75,eigen_79,durrett_tpb98,johnson_tree02,lion_el07}. At a larger spatial scale the mentioned associations can be considered as objects forming larger (super) associations and the repetition of this process can even yield a hierarchy of associations. Now some general and elementary features of this complex process are revealed by a toy model exhibiting several ways how the associations coexist.

The spatial predator-prey models with many species proved to be an appropriate model to study the formation and competition of associations \cite{szabo_pre01a,szabo_jpa05,szabo_jtb07}. In these models the associations are composed of a portion of all the species and are characterized by a spatio-temporal pattern. In fact, the associations are possible solutions and some of these solutions can be observed as a final state when the numerical simulations are performed on small systems. As the solutions of any subsystem (where several species are missing) are also solutions for the whole system therefore the number of solutions
(possible associations) increases exponentially with the number of species (excepting for some particular food webs). In some cases, in spite of the large number of possible solutions, the evolutionary process selects one of the possible solutions characterizing the final stationary state even for an infinitely large system size. In other cases, equivalent associations compete for territories through a domain growing process, as it happens for the $q$-state Potts model below the critical temperature \cite{wu_rmp82}, and finally one of the associations will prevail in the whole (finite) system. Within a wide range of parameters, however, the domain growing process is stopped and one can observe a self-organizing domain structure (sustaining all species alive) where large domains of associations can be clearly identified. The self-organizing pattern can be maintained by cyclic dominance between the associations or by other dynamical phenomena (sometimes resembling the death and rebirth of the Phoenix bird) where different length- and time-scales emerge (for examples see the Refs. \cite{szabo_jpa05,szabo_jtb07}).

Now we describe another mechanism yielding a self-organizing pattern with five associations representing two basically different classes of the defensive alliances which can be considered as privileged associations. This effect is observed in a six-species predator-prey model which is a simplified combination of two previously investigated models 
\cite{szabo_jpa05,perc_pre07b}.

\section{THE MODEL}
\label{sec:model}

\begin{figure}
\centerline{\epsfig{file=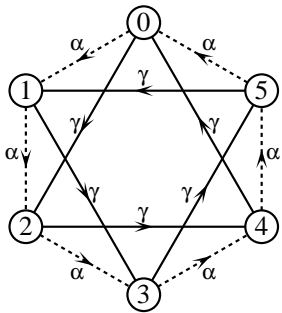,width=4cm}}
\caption{\label{fig:fw}Food web for the present six-species predator-prey model. Arrows point from a predator towards its prey with heterogeneous invasion rates specified along the edges.} 
\end{figure}

We consider a six-species predator-prey model where each site $i$ of a square lattice is occupied by an individual belonging to one of the six species. The species distribution is characterized by the set of site variables ($s_i=0, \ldots , 5$) referring to the label of species at the given site $i$. The predator-prey relations are defined by a food web indicating that each species has two predators and two prey. For the present model we distinguish two invasion rates, $\alpha$ and $\gamma$ ($0 \le \alpha, \gamma \le 1$), as demonstrated in Fig.~\ref{fig:fw}. The different values of $\alpha$ and $\gamma$ parameters describe the cases when the strengths of dominance within a cyclic alliance and between the members of different alliances are unequal.

The evolution of species distribution is controlled by repeating the following elementary steps. First, two neighboring sites ($i$ and $j$) are chosen at random. If $s_i$ is the predator of $s_j$ then the site $j$ will be occupied by an offspring of the species $s_i$ (in short, $s_j \to s_i$) with a probability given by the corresponding invasion rate ($\alpha$ or $\gamma$). Evidently, for the opposite predator-prey relation  $s_i$ will be transformed to the state $s_j$ ($s_i \to s_j$) with the corresponding
probability. If $s_i$ and $s_j$ are a neutral pair (e.g., $s_i=0$ and $s_j=3$) then they exchange their sites [$(s_i,s_j) \to (s_j,s_i)$] with a probability $X$ characterizing the strength of mixing. Finally, nothing happens if $s_i=s_j$.

The system is started from a random initial state where each species is present with the same probability. After many repetition of the above elementary steps the system evolves into a state that can be characterized by the average densities $\rho_s$ ($s=0, \ldots , 5$) satisfying the condition $\sum_{s=0}^{5} \rho_s=1$. For many cases the quantification of the nearest-neighbor pair correlations is necessary to give an adequate description about the spatial distribution of species. Therefore, we can introduce four types of pair configuration probabilities for the present model: $p_{id}$ denotes the probability of finding identical species on two neighboring sites; $p_n$ is the probability of finding a neutral pair (e.g., species 0 and 3); $p_{\alpha}$ and $p_{\gamma}$ are the sum of those predator-prey pair probabilities where the invasion rates are $\alpha$ and $\gamma$, respectively. These quantities are also satisfying a normalization condition, i.e., $p_{id} + p_n + p_{\alpha} + p_{\gamma} = 1$.

The above system was investigated by Monte Carlo (MC) simulations on a square lattice of size $N = L \times L$ under periodic boundary conditions and the linear size $L$ is varied from 400 to 4000. The MC simulations were performed systematically for a fixed value of the highest invasion rate (e.g., for $\gamma =1$) while the other invasion rate and $X$ are varied gradually. The stationary states were characterized by the above mentioned order parameters averaged over a sampling $t_s$ after a suitable thermalization time $t_{th}$. To observe the actual spatio-temporal pattern at a specified values of $\alpha$, $\gamma$, and $X$, the parameters $L$, $t_s$, and $t_{th}$ were adjusted as specified below.

Some features of this model has already been discussed previously \cite{szabo_jpa05,perc_pre07b,szabo_pre07}. The relevant solutions remain valid even for $\alpha \ne \gamma$. These solutions are the six homogeneous distributions, the two cyclic defensive alliances, and three well mixed phases of two neutral species. For the cyclic defensive alliances the odd (or even) label species invade cyclically each other in the same way as it is described by the spatial Rock-Scissors-Paper game
\cite{tainaka_prl89,szabo_pr07,nishiuchi_pa08}. The distinguished role (and also the name) of the cyclic defensive alliances come from the fact
that the self-organizing spatio-temporal pattern provides a protection (stability) against external invaders \cite{boerlijst_pd91,szabo_pre01a,he_ijmpc05}.

When $X$ is increased for $\alpha=\gamma=1$ the present system exhibits a first-order phase transition at $X=X_{c}(\alpha=\gamma=1)=0.05592(1)$ \cite{szabo_jpa05}. If $X<X_{c}(\alpha=\gamma=1)$ then one of the two cyclic defensive alliances will prevail the whole system after a domain growing process. Henceforth this final state will be denoted by $T^C$
referring to cyclic triplets. This model has three other defensive alliances composed from a neutral pair of species (e.g., 0 and 3) because in their well-mixed phase the participants protect each other mutually against any external invaders \cite{szabo_jpa05}. The MC simulations have justified that one of these two-species defensive alliances will dominate the whole lattice after a domain growing process if $X > X_c$. This latter
final state is named in short as $D$ (duet). Notice that only a portion of the species remains alive in this system for the uniform invasion rates.

The internal symmetry of the two cyclic defensive alliances is conserved in the systems for the alliance-specific invasion rates \cite{perc_pre07b}. In the latter case four different invasion rates ($\alpha$, $\beta$, $\gamma$, and $\delta$) has been distinguished on the same food web plotted in Fig.~\ref{fig:fw}. For $X=0$, this type of parametrization has allowed us to study the cases where one of the cyclic defensive alliances is preferred to the other. It turned out, for example, that the protection mechanism is enforced if the invasion rates are increased within a cyclic three-state alliance. This four-parameter model becomes equivalent to the present model for $\alpha=\beta$ and $\gamma=
\delta$ in the absence of mixing.

For the case of $\alpha=1$ and $\gamma=0$ the food web has only one (six-species) cycle. This system was already investigated previously by several authors \cite{frachebourg_jpa98,szabo_pre07}. In analogy to the spatial Rock-Scissors-Paper games the species alternates cyclically at
each site and a self-organizing pattern is maintained by the moving invasion fronts for $X=0$.

For strong mixing the formation of well-mixed phases of the neutral species is expected. The three two-species associations are equivalent for $\alpha = \gamma$ and the motion of interfaces separating them is controlled by the curvature and random events \cite{bray_ap94}. This means that if two different domains are separated by a straight-line boundary then the average velocity of this interface is zero. However, if $\alpha
> \gamma$ then the well mixed phase of species 0 and 3 can invade the territory of the well-mixed phase of species 1 and 4, that can also invade the third association (consisting of species 2 and 5). In other words, the present model exemplifies a system where three associations play the spatial Rock-Scissors-Paper game. In the opposite case ($\alpha < \gamma$) the direction of cyclic dominance is reversed. When visualizing the evolution of species distribution in this phase, one can recognize rotating spiral arms reported for many other systems
(for nice snapshots see the papers
\cite{boerlijst_pd91,durrett_tpb98,kerr_n02,hauert_s02,reichenbach_n07} and
further references therein). This phase is denoted by $T^C(D)$ signaling the
cyclic dominance of duets. Originally the recent research was planned to explore this phenomenon. It turned out, however, that the present model exhibits other self-organizing patterns as it will be detailed in the next section.

\section{THE RESULTS}
\label{sec:results}

Without loosing generality we discuss separately the cases $\alpha < \gamma$ (at $\gamma=1$) and $\gamma < \alpha=1$.

\subsection{The region $\alpha < \gamma$}
\label{sec:a<g}

\begin{figure}[ht]
\centerline{\epsfig{file=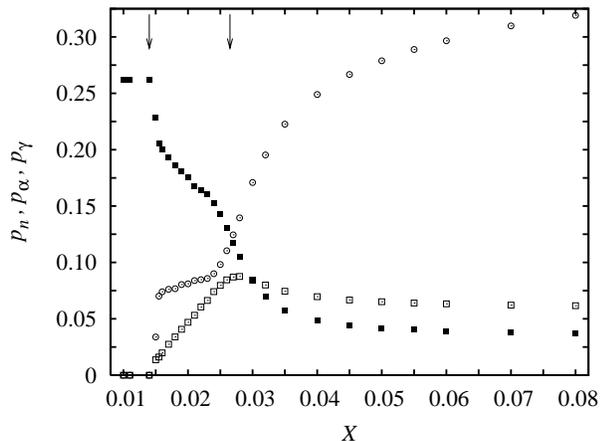,width=8cm}}
\caption{\label{fig:a04} The pair configuration probabilities $p_\alpha$ (open squares), $p_\gamma$ (closed squares) and $p_n$ (open circles) as a function of $X$ at fixed $\alpha = 0.4$ and $\gamma=1$ values. Arrows show the positions of phase transitions.}
\end{figure}

First we study MC results obtained when varying $X$ for $\alpha=0.4$ and $\gamma=1$. As previously discussed, the variation in the spatial distribution can be quantified by the above mentioned pair correlation functions, namely, $p_n$, $p_{\alpha}$, and $p_{\gamma}$. In the numerical results plotted in Fig.~\ref{fig:a04} two arrows indicate the threshold values of the mixing ($X_{c1}(\alpha)$ and $X_{c2}(\alpha)$) where phase transitions occur. 

If $X < X_{c1}(\alpha)$ then the finite system evolves into one of the $T^C$
phases after a suitable relaxation (domain growth) time increasing with
$N$. Within this phase the odd (or even) label species form a cyclic defensive
alliance where the three mentioned species are present with the same average density ($1/3$) and $p_n=p_{\alpha}=0$.

If $X > X_{c2}(\alpha)$ then the three well mixed associations of the neutral
pairs form a self-organizing pattern ($T^C(D)$) resembling to the spatial Rock-Scissors-Paper game at a higher level. The typical extensions of domains and the width of boundary layers (separating two associations of neutral pairs) depend on $X$ and $\alpha$. The qualitative analysis indicates an increase in the typical domain size if $\alpha$ goes to $\gamma=1$ [providing $X>X_{c1}(\alpha=\gamma=1)=X_{c2}(\alpha=\gamma=1)=0.05592(1)$]. In fact, the driving force of the cyclic dominance is proportional to $\gamma - \alpha$. The numerical study of the impact of the vanishing cyclic dominance on the spatial distributions was already presented in a model combining the three-state Potts model and spatial Rock-Scissors-Paper game \cite{szabo_pre02a,szolnoki_pre05a}. In the light of the latter results it is expected that the typical domain size increases as $1/|\alpha - \gamma|$ when approaching the symmetric case ($\alpha=\gamma$).

\begin{figure}[ht]
\centerline{\epsfig{file=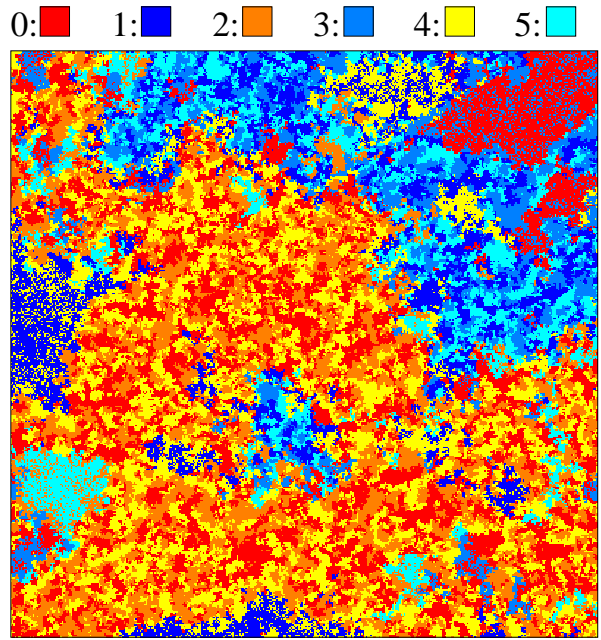,width=8cm}}
\caption{\label{fig:snapshot}(Color online) Typical spatial distribution of species within a box of size $400 \times 400$ for $\alpha=0.4$, $\gamma = 1$, and $X=0.02$.}
\end{figure}

The appearance of an intermediate region [$X_{c1}(\alpha) < X < X_{c2}(\alpha)$] in Fig.~\ref{fig:a04} was unexpected. The visualization of the evolution of species distribution (for a snapshot see Fig.\ref{fig:snapshot}) have indicated clearly that within this parameter region five types of domains (associations) can be distinguished. Namely, the two cyclic triplets ($T$) and also the three associations of neutral pairs ($D$) which form a self-organizing domain structure.

It is worth emphasizing that a sufficiently large system size and long runs were necessary in the MC simulations to observe this intermediate region. More precisely, the self-organizing patterns has reached their final features (domain size, {\it etc}.) after a typical time of $t_{th} = 4 \times 10^5$ MCSs if $L=4000$. For the sake of comparison, the quantitative features of the $T^C(D)$ pattern can be well studied for $L=400$ after $t_{th} = 4 \times 10^4$ MCSs.

Previous analyzes of similar systems have justified that the value of the critical point can also be determined by evaluating the average velocity of a straight invasion front separating two phases characterizing the final behavior below and above the critical point. The average velocity of this invasion front becomes zero at the critical point. To clarify the behaviors in the intermediate region we have performed these numerical investigations for different values of $X$. The results have clearly indicated that each $D$ state can invade the territory of the $T$ associations within the intermediate state. In other words, if an island of $D$ (with a sufficiently large extension) is created via a nucleation mechanism within the territory of $T$ (or even at the boundary of two $T$ states) then this island grows permanently. 

\begin{figure}[ht]
\centerline{\epsfig{file=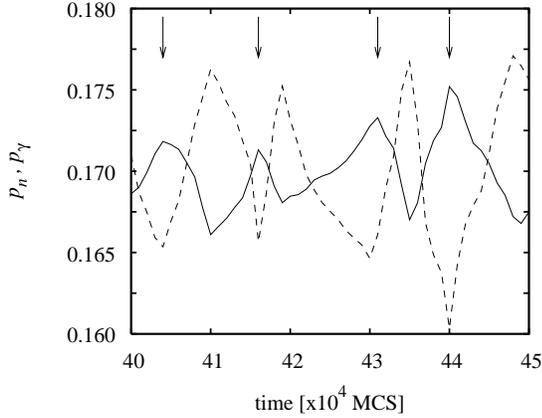,width=8cm}}
\caption{\label{fig:evol} Variation of pair configuration probabilities
  [$p_\gamma(t)$ (solid) and $p_n(t)$ (dashed line)] in the stationary state
  within the intermediate region ($\alpha=0.4$, $\gamma=1$, and $X=0.021$) for
  $L=4000$. The values of $p_n$ are increased by a constant for easier
  comparison. Arrows show when $D$ domains start to expand.}
\end{figure}

In the present case, however, three equivalent $D$ associations exist which dominate cyclically each other as mentioned above. Consequently, within the intermediate phase two growing $D$ phases can collide and one of them will invade the other. During the invasions the moving invasion front leave behind a slowly varying structure that differs from its final (well-mixed) distribution. Thus, the expanding $D$ associations become less stable against the invasion of the neighboring $T$ associations in the vicinity of the moving invasion fronts. The visualization of the species distribution has demonstrated clearly that in many cases the newly invaded $D$ territories were occupied by the neighboring $T$ associations within a transient time. The alternative expansion of $D$ and $T$ domains can be traced well by monitoring the evolution of pair configuration probabilities. Evidently, the growth of $D$ domains involves the increase of $p_n$ while the extension of $T$ domains increases the average value of $p_{\gamma}$. Consequently, one can observe opposite variations in the time-dependence of $p_n(t)$ and $p_{\gamma}(t)$ as demonstrated in Fig.~\ref{fig:evol}. Evidently, the "amplitude" of these variations vanish in the limit $N \to \infty$.

Increasing $X$ yields faster recovering (shorter transient time) and simultaneously makes the $D$ territories more stable against the invasion of $T$ domains. As a result, above a second threshold value ($X > X_{c2}(\alpha)$) the $T$ associations cannot remain alive and the whole system is prevailed by the previously described $T^C(D)$ phase. 

\begin{figure}[ht]
\centerline{\epsfig{file=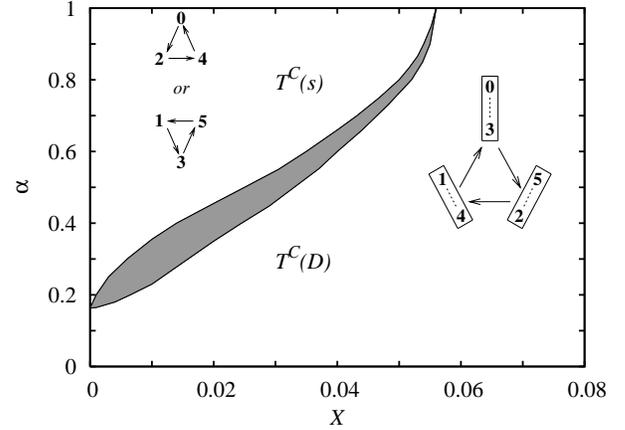,width=8cm}}
\caption{\label{fig:phda} Phase diagram of the model as a function of $X$ and $\alpha$ for $\gamma=1$. The $T^C(s)$ region is characterized by the exclusive dominance of ($0+2+4$) or ($1+3+5$) cyclic alliance. The area $T^C(D)$ corresponds to the phase where three alliances of two-neutral species [($0+3$), ($1+4$), and ($2+5$)] play spatial Rock-Scissors-Paper game. Within the shadowed region all the mentioned associations coexist and form a self-organizing pattern described in the text.}
\end{figure}

In order to determine the critical values of mixing ($X_{c1}(\alpha)$ and $X_{c2}(\alpha)$), the MC simulations were repeated  with increasing gradually the value of $X$ for several values of $\alpha$. The results are summarized in a phase diagram shown in Fig.~\ref{fig:phda}.

The intermediate region vanishes if $\alpha < \alpha_{c}=0.170(5)$. More precisely, $X_{c1}(\alpha)$ and $ X_{c2}(\alpha)$ go to zero simultaneously if $\alpha$ tends to $\alpha_{c}$ from above. For $\alpha < \alpha_{c}$ the $T^C(D)$ state occurs after a relaxation proportional to $1/X$ in the limit $X \to 0$. In the absence of local mixing ($X=0$), however, the well-mixed state of the neutral pairs cannot occur and the system develops into a state where the evolution of species distribution is governed by invasions of type $\gamma$ in a pattern exhibiting large domains of $T$ associations.

\subsection{The region $\gamma < \alpha$}
\label{sec:g<a}

\begin{figure}[ht]
\centerline{\epsfig{file=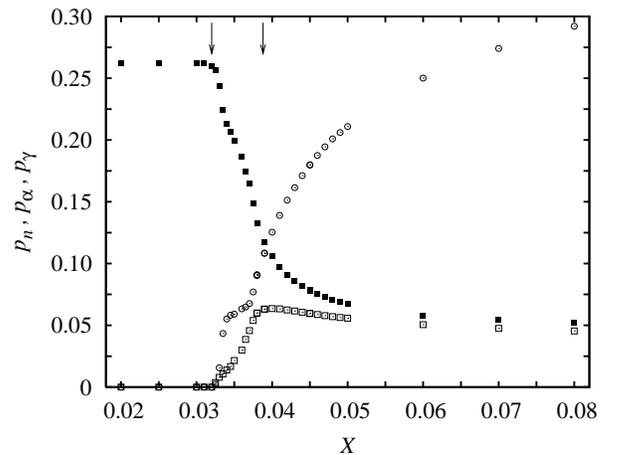,width=8cm}}
\caption{\label{fig:g06} The pair configuration probabilities, such as $p_\alpha$ (open squares), $p_\gamma$ (closed squares) and $p_n$ (open circles) as a function of $X$ at fixed $\gamma = 0.6$ and $\alpha=1$ values. Arrows point to the positions of phase transitions.}
\end{figure}

Similarly to the previous section now we study the case of $\gamma < \alpha=1$. Within this range of parameters the direction of cyclic invasions between the well mixed alliances of neutral species pair is reversed, that is, ($0,3$) dominates ($1,4)$ dominates ($2,5$) dominates ($0,3$). As the direction of cyclic invasions [within the phase of $T^C(D)$] do not affect the the main features of pattern formation therefore we expect a phase diagram similar to the case of $\alpha < \gamma$. This expectation is supported by Fig.~\ref{fig:g06} where the $X$ dependences of pair configuration probabilities are plotted for $\gamma=0.6$. The qualitative similarity between Figs.~\ref{fig:a04} and \ref{fig:g06} is striking.

\begin{figure}[ht]
\centerline{\epsfig{file=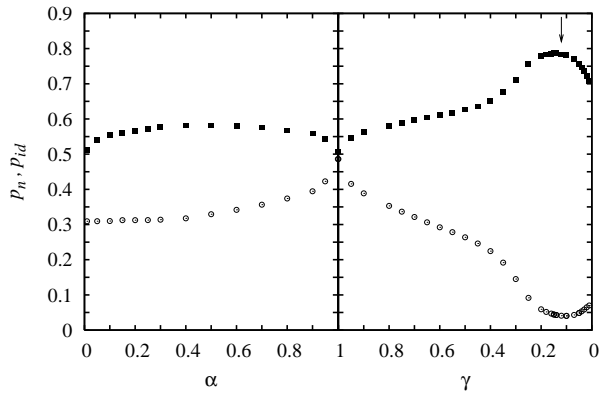,width=8cm}}
\caption{\label{fig:x08} The $\alpha$ and $\gamma$ dependence of $p_n$ (open circles) and $p_{id}$ (closed squares) at a fixed value of mixing ($X= 0.08$). The arrow indicates the minimum of $p_n$.}
\end{figure}

Figure \ref{fig:g06} represents a situation where both types ($\alpha$ and $\gamma$) of invasions play a relevant role in the pattern formation. In the opposite case (when either $\alpha$ or $\gamma$ tends to zero) a relevant difference occurs in the behaviors as plotted in Fig.~\ref{fig:x08} comparing $p_n$ changes as a function of $\alpha$ and $\gamma$ for a fixed value of $X$. 

Figure~\ref{fig:x08} shows that $p_n$ has a local minimum at a small value of $\gamma$ which is missing in the case of $\alpha<\gamma$. In fact, for $\gamma = 0$ the system develops into a state [denoted as $S^C(s)$ (cyclic sextet)] where the six species invades cyclically each other. Within this spatio-temporal pattern the site-exchange process becomes rare and cannot affect significantly the spatial distribution of species. On the other hand, we can observe a smooth transition from the state $T^C(D)$ to $S^C(s)$ which is accompanied with the suppression of $D$ domains (yielding a relevant decrease in $p_n$) and with an increase of $p_{\alpha}$ and $p_{id}$ when decreasing the ratio $\gamma / \alpha$. All these processes together result in a minimum in $p_n$ that used to define a phase boundary between these phases. 

\begin{figure}[ht]
\centerline{\epsfig{file=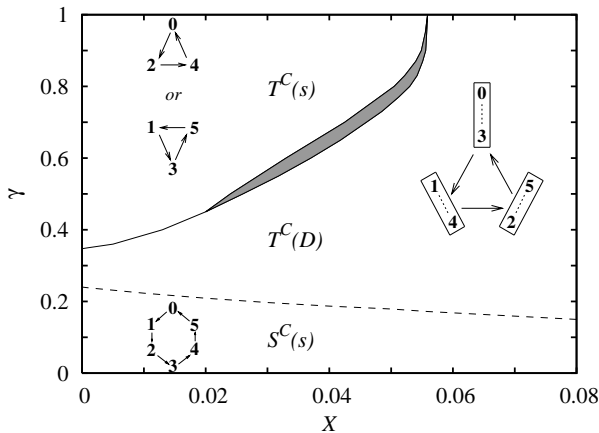,width=8cm}}
\caption{\label{fig:phdg} Phase diagram of the model as a function of $X$ and $\gamma$ for $\alpha=1$. The notation of phases are the same as for Fig.\ref{fig:phda}. $S^C(s)$ refers to a spatio-temporal pattern in which the evolution is dominantly governed by the cyclic invasions of the six species.}
\end{figure}

The $\gamma - X$ phase diagram is shown in Fig.~\ref{fig:phdg} where the dashed line shows the value of parameters where the minimum occurs in $p_n$. According to our numerical investigations the value of $X_{c1}(\gamma)$ and $X_{c2}(\gamma)$ coincide in this phase diagram within a range of $\alpha$  ($0.35(2) < \alpha < 0.45(2)$) and both quantities vanish if $\alpha < 0.35(2)$.

\section{SUMMARY}
\label{sec:sum}

We have studied a six-species ecological model on a square lattice where different types of associations are formed from a portion of species existing in the whole ecological model. The investigation of the present model was inspired by previous results exemplifying several ways how the cyclic dominance can occur between the associations characterized by their composition and spatio-temporal pattern. In most of the previous studies the number of parameters was reduced by introducing many symmetries. Now we wished to explore some further phenomena that yield the formation and competition of alliances in more realistic biological systems when the symmetries are reduced in the invasion rates. More precisely we have studied the cases characterized by two invasion rates, $\alpha$ and $\gamma$, in a way preserving the internal symmetries of the cyclic triplets.  

The present model exhibit a wide range of behaviors in the final stationary
states as summarized in two phase diagrams (see Figs.~\ref{fig:phda} and
\ref{fig:phdg}). For example, if $\alpha \ne \gamma$ and the site exchange
mechanism is sufficiently strong then one observes a self-organizing
spatio-temporal pattern in which the three alliances of neutral pairs dominate
cyclically each other. Although similar self-organizing patterns are reported
in other systems \cite{szabo_jtb07}, the present one seems to be the simplest
lattice predator-prey model where the mechanism of cyclic dominance can take
place at two different levels. In addition to this feature we have also
revealed an unexpected phase where both the domains of cyclic three-species
alliances and the neutral two-species alliances can coexist. The existence of
this intermediate phase is closely related to the emergence of different time-
and length-scales within the self-organizing patterns. We think that further
reduction of the symmetries in the species specific invasion rates can yield
more complex behaviors and other uncovered mechanisms supporting 
the coexistence of different alliances of species.

\begin{acknowledgments}

This work was supported by the Hungarian National Research Fund
(Grant Nos. T-47003 and K-73449).

\end{acknowledgments}


\end{document}